\documentclass{elsart}
\usepackage{graphicx}
\usepackage{amsmath}
\usepackage{amssymb}

\begin{document}

\begin{frontmatter}
\title{Field-tunable magnetic phases in a semiconductor-based two-dimensional Kondo lattice}

\author[address1]{C. Siegert},
\author[address2]{A. Ghosh},
\author[address1]{M. Pepper},
\author[address1]{I. Farrer},
\author[address1]{D. A. Ritchie},
\author[address1]{D. Anderson}, and
\author[address1]{G. A. C. Jones}

\address[address1]{Cavendish Laboratory, University of Cambridge, J.J. Thomson Avenue, Cambridge CB3 0HE, United Kingdom.}
\address[address2]{Department of Physics, Indian Institute of Science, Bangalore 560 012, India.}

\begin{abstract}
We show the existence of intrinsic localized spins in mesoscopic
high-mobility GaAs/AlGaAs heterostructures. Non-equilibrium transport
spectroscopy reveals a quasi-regular distribution of the spins, and
indicates that the spins interact indirectly via the conduction
electrons. The interaction between spins manifests in characteristic
zero-bias anomaly near the Fermi energy, and indicates gate
voltage-controllable magnetic phases in high-mobility
heterostructures. To address this issue further, we have also
designed electrostatically tunable Hall devices, that allow a
probing of Hall characteristics at the active region of the
mesoscopic devices. We show that the zero field Hall coefficient has
an anomalous contribution, which can be attributed to scattering by
the localized spins. The anomalous contribution can be destroyed by
an increase in temperature, source drain bias, or field range.
\end{abstract}

\end{frontmatter}

\section{Introduction}

Experimental simulations of fundamental magnetic interactions on few
spin basis are very challenging. The most common approach
investigates the interaction between localized spins with the spin
cloud of surrounding conduction electrons, where localized spins are
created by using either atomic nucleii \cite{smet2002},
intentionally embedded magnetic impurities \cite{madhavan1998}, or
odd electron quantum dots (QDs)
\cite{goldhaber-gordon1998,cronenwett1998}. Individual localized
spins can be employed to simulate the Kondo effect
\cite{goldhaber-gordon1998,cronenwett1998}. In systems containing
more than one localized spin, the spins can interact via the
Ruderman-Kittel-Kasuya-Yoshida (RKKY) exchange with interaction
parameter $J^{\rm RKKY}$~\cite{jayaprakash_HRK,affleck}. This has
been shown experimentally for the two-spin case, where the
interaction has been tuned electrically with surface gates
\cite{craig2004}. The pairwise coupling can be ferromagnetic for
$J^{\rm RKKY} < 0$, or anti-ferromagnetic for $J^{\rm RKKY} > 0$.
For non-interacting spins, $J^{\rm RKKY} = 0$, the Kondo effect
contributes one additional state at each localized spin within a
bandwidth of $T_{\rm K}$ at $E_{\rm F}$. Non-equilibrium transport
spectroscopy then shows a resonance in differential conductance
${\rm d}I/{\rm d}V$ at zero source-drain bias ($V_{\rm SD} = 0$)
\cite{kouwenhoven2001,ghosh2005}, which we call type-I zero bias
anomaly (ZBA). With interacting spins $|J^{\rm RKKY}| > 0$, the
resonance is suppressed at $E_{\rm F}$ for $k_{\rm B} T < |J^{\rm
RKKY}|$, and $|V_{\rm SD}| < |J^{\rm RKKY}|$, resulting in a split
resonance, which we call type-II ZBA
\cite{heersche2006,wiel2002,ghosh2004,siegert2007}. The half width
at half maximum of the split is defined as $\Delta$ and acts as an
indicator of spontaneous spin-polarization and static magnetic
phases within the system~\cite{pasupathy,heersche2006,cobden}.

Here we present high-mobility mesoscopic heterostructures that show
signatures of an intrinsic spin lattice. As, in two dimensions,
interspin distance $R$, and $k_{\rm F} = \sqrt{2 \pi n_{\rm 2D}}$
tunes the exchange interaction parameter $J^{\rm RKKY}(k_{\rm F},R)
\propto {\rm cos}(2 k_{\rm F} R)/R^2$ for $k_{\rm F}R >> 1$
\cite{Beal-Monod1987,siegert2007}, we can tune the magnetic state of
the spin lattice with a simple non-magnetic surface gate. The
interspin distance $R$ is fixed for a given device, and can be
evaluated via phase interference effects at low temperatures, or
with commensurability effects at temperatures $T > 500$~mK \cite{siegert2007}. Such a
system is ideal for investigations of magnetic phases in
two-dimensions. As non-equilibrium transport spectroscopy shows the
magnitude $|J^{\rm RKKY}|$ only, zero field Hall measurements have
to be carried out that might enable the determination of the sign of
$J^{\rm RKKY}$.

\section{Two probe transport spectroscopy}
\begin{figure}\label{Fig1}\centering
\includegraphics[width=7cm]{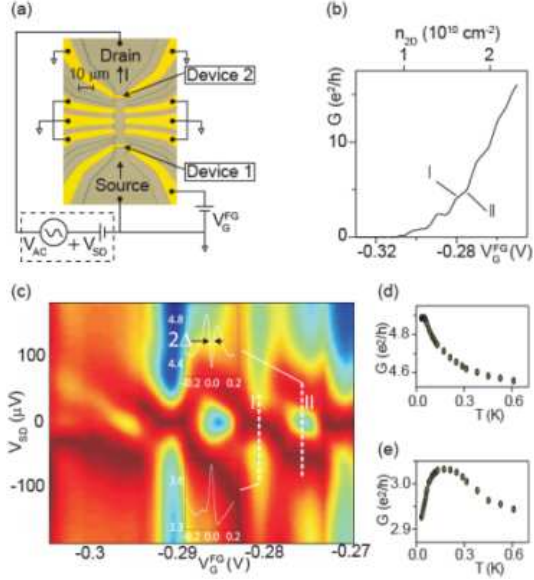}
\caption{(a) Optical image of typical device, including electrical connections. For operation, all gates are grounded with the exception of $V^{\rm FG}_{\rm G}$, which is used to vary the electron density in a $2 \times 8 \mu$m$^2$ region. Mixed AC+DC voltage is used to perform non-equilibrium transport spectroscopy. (b) At 35 mK typical linear conductance $G(V^{\rm FG}_{\rm G})$, and $n_{\rm 2D}(V^{\rm FG}_{\rm G})$ in the upper panel. (c) Surface plot d$I$/d$V$ over $V^{\rm FG}_{\rm G}$ showing alternating type-I, and type-II ZBA. In the white insets typical type-I, and type-II ZBA are shown. (d) Typical temperature dependence of type-I ZBA showing the peak vanishes monotonically over $T$. (e) Typical temperature dependence of type-II ZBA showing non-monotonicity at $V_{\rm SD} = 0$.}
\end{figure}

Devices were fabricated from high mobility Si-$\delta$-doped
GaAs/Al$_{0.33}$Ga$_{0.67}$As heterostructures. The 2DEG is formed
300 nm below the surface, and the spacer layer between dopants and
the 2DEG is 80~nm. Si doping density of $n_{\rm Si} \approx 2.5
\times 10^{12} $~cm$^{-2}$ results in as-grown electron density
$n_{\rm 2D} \approx 1 \times 10^{11}$~cm$^{-2}$ with low temperature
mobilities $\sim 1-3\times10^6$~cm$^2$/V-s. By selective wet-etching
a 8~$\mu$m wide mesa is created. One thermally deposited 2~$\mu$m
long non-magnetic Ti/Au gate on the mesa is used to
electrostatically vary $n_{\rm 2D} \approx 1-3 \times
10^{10}$~cm$^{-2}$ in the active $2 \times 8 \mu$m$^2$ region. Fig.
1(a) shows an optical microscope image of a typical device with
electric setup.

The devices were characterized in dilution refrigerators with base
temperatures down to $\approx 30$~mK. Combined AC+DC two-probe
measurements with ac excitation voltage $V_{\rm AC} \ll k_{\rm
B}T/e$ is used for non-equilibrium transport spectroscopy. In
fig.~1(b) we show a typical linear conductance $G(V^{\rm FG}_{\rm
G})$, and $n_{\rm 2D} (V^{\rm FG}_{\rm G})$ for 30~mK at zero
magnetic field. At temperatures $T < 100$ mK, $G(V^{\rm FG}_{\rm
G})$ shows characteristic features of type-I, and type-II ZBA. For
most devices, the structures are strongest in the electron density
range $n_{\rm 2D} \sim 1-3\times10^{10}$~cm$^{-2}$, see fig.~1(b),
and often visible up to $G \gtrsim 10-15\times(e^2/h)$. In
non-equilibrium transport spectroscopy d$I$/d$V (V^{\rm FG}_{\rm G},
V_{\rm SD})$ consists of a repetitive sequence of two-types of
resonances as $V_{\rm G}$ is increased, see fig.~1(c). The
single-peaked type-I ZBA splits intermittently to form a gap at
$E_{\rm F}$, resulting in type-II ZBA. The insets (white) show the
non-equilibrium traces of type-I ZBA, and type-II ZBA at the $V^{\rm
FG}_{\rm G}$ indicated in fig.~1(b).

Type-I ZBA decrease monotonically with increasing T, see fig.~1(d),
and with increasing parallel magnetic field B \cite{ghosh2005}. In
non-equilibrium transport spectroscopy type-I ZBA split with
$2g\mu_{\rm B}B$ \cite{siegert2007,ghosh2005}. Type-II ZBA show a
non-monotonicity at the scale of $\Delta$ over both T (see
fig.~1(e)), and B, which is characteristic for interacting spins
$J^{\rm RKKY} > 0$ \cite{heersche2006,siegert2007,ghosh2005}. Type-I
ZBA can then be understood as a spin lattice with $|J^{\rm RKKY}|
\ll k_{\rm B}T$. While the existence of mutually interacting
localized spins in high-mobility mesoscopic heterostructures can
account for such behavior, the question of spatial arrangement of
the spins needs to be addressed independently with
magnetoconductance measurements.

\begin{figure}\label{Fig2}\centering
\includegraphics[width=7cm]{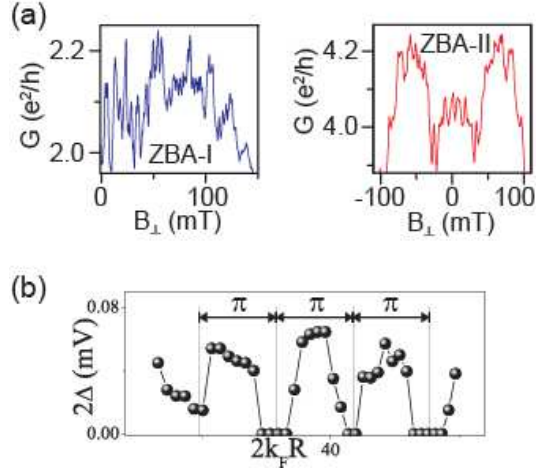}
\caption{(a) Low temperature $G(B_\bot)$ for type-I, and type-II ZBA showing Aharonov-Bohm like oscillations, which allow the determination of a regular lattice parameter R. (b) Using R from AB-like oscillations, and $n_{\rm 2D}$ from capacitance model, $\Delta$ is found to be periodic in $\pi$ over $2k_{\rm F}R$.}
\end{figure}

In two-dimensional spin lattice models, with localized spins arising
from fluctuations in the conduction band, both Aharonov-Bohm like
oscillations, as well as commensurability effects are expected
\cite{siegert2007}. Sites of localizes spins act as scattering
centers, or as antidot, for conduction electrons, leading to
back-scattering and interference effects when the phase coherence
length of electrons exceeds $R$. At low $T$, a random spatial
distribution of the scatterers or antidots gives rise to weak
localization phenomena. The magnetoconductance $G(B_\bot)$ then
shows universal aperiodic fluctuations. However, for a quasi-regular
antidot array, transport can be modeled through multiple, connected
Aharonov-Bohm rings encircling a discrete number of impurities.
Aharonov-Bohm like oscillations with periodicity $\Delta B_\perp
\approx h/\pi eR_{\rm C}^2$, where $R_{\rm C}$ is the radius of a
stable ring~\cite{schuster1994,weiss}, appear. The oscillations can
be assigned to the respective orbits via Fourier Transform, and the
lowest possible orbit gives an estimate for $R$~\cite{siegert2007}.
Fig.~2(a) shows typical oscillations in the magnetoconductance for a
type-I, and a type-II ZBA. We cross-checked the quasi regular
distribution of the localized spins at higher $T$ from
commensurability effects of classical electron trajectories in a 2D
array of point scatterers at finite $B_\bot$~\cite{weiss1,siegert2007}. Please note, that the magnetoconductance oscillations appear only over a range of $n_{\rm 2D}$, where ZBA is observed.

Associating each antidot to one (or more) localized spin then
realizes a spin array embedded in the sea of conduction
electrons~\cite{siegert2007}. These spins can interact via the
conduction electrons by RKKY mechanism, and are then expected to
have a characteristic periodicity of $\delta (2k_{\rm F}R) = \pi$ in
$|J^{\rm RKKY}|$. As $\Delta \propto |J^{\rm RKKY}|$, we can confirm
this by plotting in fig.~2(b) $\Delta (2k_{\rm F}R)$. The clear
periodicity of $\approx \pi$ is recognized as the for RKKY typical
$2k_{\rm F}R$-oscillations, establishing the existence of an
interacting spin lattice in high mobility heterostructures
\cite{siegert2007}.

\section{Zero field Hall coefficient}
\begin{figure}\label{Fig3}\centering
\includegraphics[width=7cm]{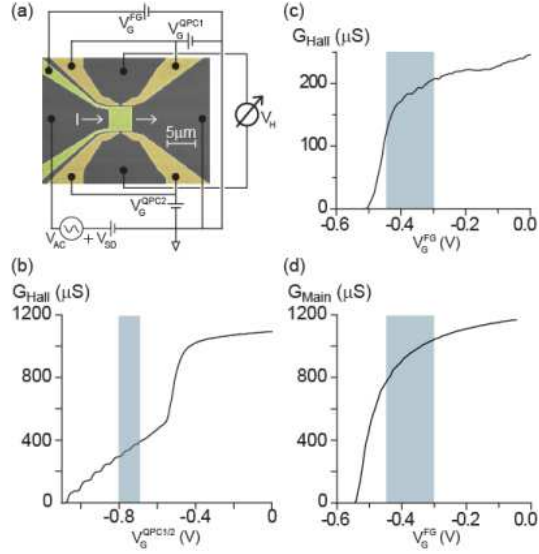}
\caption{(a) Optical image of a $4 \times 4 \mu$m$^2$ Hall device. $V^{\rm QPC1}_{\rm G}$, and $V^{\rm QPC1}_{\rm G}$ define a 4 $\mu$m wide mesa with QPC to probe Hall voltage. $V^{\rm FG}_{\rm G}$ varies $n_{\rm 2D}$ in the active $4 \times 4 \mu$m$^2$ region. (b) Voltages $V^{\rm QPC1/2}_{\rm G}$ are operated between -0.8V and -0.7V to ensure open channel with strong confinement. (c) For both QPC voltages $V^{\rm QPC1/2}_{\rm G} = -0.8V$, the full gate $V^{\rm FG}_{\rm G}$ is operated such that the conductance in Hall direction does not drop below $2e^2/h$ (grayish region). (d) For same conditions conduction in the main direction, as used for the experiment.}
\end{figure}

While non-equilibrium transport spectroscopy provides an indirect
evidence of localized spins in high-mobility heterostructures, a
convenient, and more direct, mean to probe the magnetic state of a
two-dimensional spin lattice is the initial, or zero field, Hall
coefficient. In case of magnetic order, the finite magnetization of
the system adds an additional term to the Hall constant, resulting
in the anomalous Hall effect (AHE) \cite{Toyosaki2004}. The Hall
resistance is then $R_{\rm H} = \gamma_{\rm H}^0 B_\bot +
\gamma_{\rm A} M$, with $B_\bot$ being magnetic field, M being
magnetization, $\gamma_{\rm H}^0 = -1/n_{\rm 2D}|e|$ in two
dimensions, and $\gamma_{\rm A}$ being the anomalous Hall constant.
We have designed fully tunable Hall devices, and show in fig.~3(a)
an optical image with electrical setup. The 2DEG is laterally
confined by two pairs of nonmagnetic Ti/Au sidegates with voltages
V$_{\rm G}^{\rm QPC1}$ and V$_{\rm G}^{\rm QPC2}$ between -0.7~V and
-0.8~V, which ensures defined QPCs, combined with low resistance
Hall probes. Fig.~3(b) shows typical $G_{\rm Hall}(V^{\rm QPC}_{\rm
G})$ for the QPCs. The voltages V$_{\rm G}^{\rm QPC1}$ and V$_{\rm
G}^{\rm QPC2}$ are adjusted at the beginning of the measurement, and
kept constant during the whole experiment. The resulting 4-5 $\mu$m
wide electrostatically defined mesa is patterned with a full Ti/Au
gate of dimensions $4\times 4$ $\mu$m$^2$ or $5 \times 5$ $\mu$m$^2$
in between the two side gates. Voltage on the full gate V$_{\rm
G}^{\rm FG}$ is applied to vary $E_{\rm F}$ and thus create the
active part of the device, with n$_{\rm 2D} = 2 - 5 \times 10^{10}$
cm$^{-2}$. In fig.~3(c) we show $G_{\rm Hall}(V^{\rm FG}_{\rm G})$
for -0.8~V at the QPCs, and the range in which $V^{\rm FG}_{\rm G}$
is operated in gray, which is chosen to ensure G$_{\rm Hall} \gg 2
\frac{e^2}{h}$ for all V$_{\rm G}^{\rm FG}$ used.

With V$_{\rm G}^{\rm QPC1}$ and V$_{\rm G}^{\rm QPC2}$ both set at
$-0.8$~V, the conductance $G$ in the longitudinal direction was
maintained at $G > 10 e^2/h$ for all $V^{\rm FG}_{\rm G}$ when the
device was in operation for the Hall measurements. Please note that
n$_{\rm 2D}$ and $G$ are higher in this experiment than in earlier
ones to ensure free paths both in longitudinal, as well as in Hall
direction. The electron density is determined at high n$_{\rm 2D}$
in the range of $\pm$ 50 mT, where no fluctuations are visible, and
both low, and high T result in the same electron density (see inset of fig.~5). Lower electron densities were determined
by extrapolating the dependence of $n_{\rm 2D}$ on $V_{\rm G}$ at
the large  $n_{\rm 2D}$ regime via standard capacitance model.

\begin{figure}\label{Fig4}\centering
\includegraphics[width=7cm]{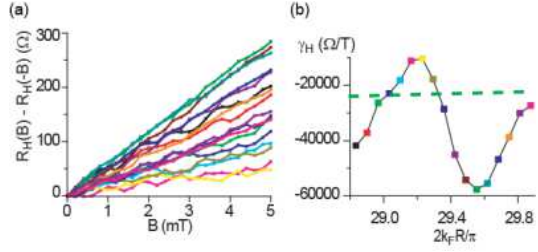}
\caption{(a) Asymmetric Hall resistance R$_{\rm H} = \frac{1}{2 I} (V_{\rm H} (B_\bot) - V_{\rm H} (- B_\bot))$ up to 5~mT, with different colors indicating different electron densities. (b) The respective Hall coefficients $\gamma_{\rm H}$ over $2k_{\rm F}R/\pi$ showing a strong oscillation. The high temperature, undistorted $\gamma_{\rm H}$ is shown as dashed green line.}
\end{figure}

The samples are measured in a dilution refrigerator at base electron
temperature $T \approx$ 75 mK. Two probe low frequency ($\approx$ 13
Hz) transport measurement with AC voltage V$_{\rm AC} \ll k_{\rm
B}T/e$ , and source-drain bias V$_{\rm SD}$ is carried out such that
current flows in the main direction through the active part only.
Perpendicular magnetic field B$_\bot$ is applied to measure the
initial Hall coefficient $\gamma_{\rm H}$, taken from the slope of
the asymmetric Hall resistance R$_{\rm H} = \frac{1}{2 I} (V_{\rm H}
(B_\bot) - V_{\rm H} (- B_\bot))$, with $I$ being the current, and
$V_{\rm H} (B_\bot)$ being the Hall voltage on the QPCs, over the
initial 2-5~mT. The asymmetric Hall resistance is taken only, to
cancel out misalignment of Hall probes, and other field symmetric
artifacts \cite{teizer2003}. In fig.~4(a) we show the asymmetric
component of the Hall resistance up to 5~mT, with different colors
belonging to different electron densities. Via Aharonov-Bohm-type
oscillations (Fig.~2) we find the interspin distance R $\approx 1.1
\mu$m in these devices. The slope of the Hall resistance gives the
Hall coefficient, which we show for the respective electron
densities in fig.~4(b). The dashed green line indicates the
unperturbed value of $\gamma_{\rm H}$, and $\gamma_{\rm H}$ shows a strong deviation
from its high temperature value.

\begin{figure}\label{Fig5}\centering
\includegraphics[width=7cm]{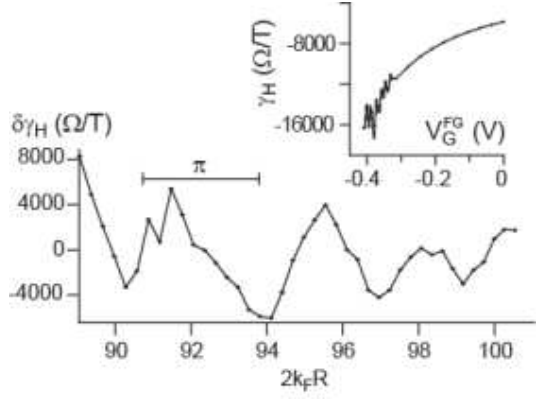}\caption{The deviation $\delta \gamma_{\rm H} = \gamma_{\rm H} - \gamma_{\rm H}^0$ over a larger range of $2k_{\rm F} R$ shows periodic fluctuations with both positive, and negative sign, and periodicity of $\pi$. Inset: Hall coefficient at high density, and high field range, as used for calculation of electron density.}
\end{figure}

The Hall coefficient deviates both positively and negatively with
respect to the unperturbed value $\gamma_{\rm H}^0$. The positive
contribution to $\gamma_H$ is unexpected, and has the wrong sign for
a magnetic contribution via the anomalous Hall effect. The negative
deviation is expected for a magnetic interaction in a spin lattice.
To investigate that further, we have considered a larger range of
densities, and show in fig.~5 the range $2k_{\rm F}R = 89 - 101$.
For clarity, only the deviation $\delta \gamma_{\rm H} = \gamma_{\rm
H} - \gamma_{\rm H}^0$ is plotted. For a given period in $2k_{\rm
F}R$, the deviation $\delta \gamma_{\rm H}$ again is either positive
or negative, and overall, the magnitude of this deviation decreases
when the carrier density is increased. This can be due to the
screening of localized spins, which increases with increasing
electron density, and reduces the visibility of the localized spins.
$\delta \gamma_{\rm H}$ has a rough periodicity of $\delta (2k_{\rm
F}R)$ = $\pi$, indicating a common origin of these fluctuations and
the ZBA in nonequilibrium transport.

Please note that clean periodicity of fluctuations in $\gamma_{\rm
H}$ requires the whole device area to be in the same phase. Since
the formation of the intrinsic spin lattices does not yield perfect
regularity in interspin distance $R$ over the whole $4\times 4$
$\mu$m$^2$ or $5 \times 5$ $\mu$m$^2$ in 100\% of our devices, we do
also observe random fluctuations, and show in fig.~6c) also a bad device.

\subsection{Stability of fluctuations}
\begin{figure}\label{Fig6}\centering
\includegraphics[width=5cm]{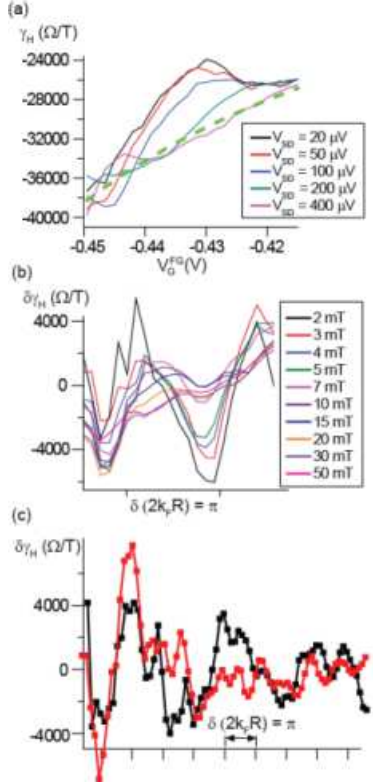}\caption{(a) Non-equilibrium $\gamma_{\rm H}$, where application of $V_{\rm SD}$ destroys the fluctuations in $\gamma_{\rm H}$. The dashed green line indicates the extrapolated $\gamma_{\rm H}$ from capacitance model. (b) $\delta \gamma_{\rm H}$ taken with various field ranges. The fluctuations in $\gamma_{\rm H}$ vanish for higher field ranges. (See text) (c) $\delta \gamma_{\rm H}$ at low $n_{\rm 2D}$ showing the effect of disorder, and only few regular fluctuations. The black trace is for confinement voltage of -0.8~V, and the red trace is taken with -0.7~V.}
\end{figure}

In fig.~6(a) we show a typical fluctuation of $\gamma_{\rm H}$ for
various source drain biases. For all other experiments $V_{\rm SD} =
0$ was kept constant. The fluctuation vanishes with increasing
$V_{\rm SD}$, and is back to its unperturbed value $\gamma_{\rm
H}^0$ (shown as dashed green line) at about $V_{\rm SD} \approx 400
\mu V$.

In fig.~6(b) we show the deviation of $\gamma_{\rm H}$, calculated
from slope of the asymmetric Hall resistance in various B field
ranges, from $\gamma_{\rm H}^0$. The lowest field ranges show
strongest fluctuations, and increasing field range decreases the
intensity of the fluctuations. Since the interspin interaction
itself couples to very small transverse fields \cite{siegert2007-2}, the fluctuations are only meaningful when
the slope is computed over smallest field ranges, i.e., leading to
the zero field Hall coefficient. Since very low field ranges do also
increase the error in the data, we use 2-5~mT range for all
$\gamma_{\rm H}$ data. We find that the highest field range 50~mT is
mostly similar to the unperturbed Hall coefficient evaluated from
the high density range and capacitance model.

By increasing the temperature the fluctuations in $\gamma_{\rm H}$
vanish (not shown), confirming the oscillations in Hall coefficient
to be a low $T$ effect.

As mentioned before, disorder appears to be a crucial aspect of the
fluctuations in $\gamma_{\rm H}$. We therefore show in fig.~6(c) an
example of a bad sample with strong disorder, and mostly irregular
fluctuations in $\delta \gamma_{\rm H}$. While no periodicity over
the whole range can be seen, most adjacent fluctuations are still
showing a rough $\pi$-periodicity, although not as clear as the less
disordered devices. While the overall magnitude of fluctuations is
increasing with lower densities, we find that the clean regularity
of the fluctuations is decreasing with decreasing $n_{\rm 2D}$,
where more random scattering appears. To avoid this we now record
data at higher electron densities only, and find much cleaner
$\pi$-periodic fluctuations, with only very little disorder.

The role of the confinement voltage is also shown in fig.~6(c). The
red trace is shown for a confinement voltage of -0.7~V, and the
black one for -0.8~V. The fluctuations in $\gamma_{\rm H}$ change
the overall appearance with change in gate voltage, but not the
characteristics. This is a very important point to note, as the
state of the device should only weakly depend on the confinement
voltage. When comparing the red and black traces, then minima, and
maxima of $\delta \gamma_{\rm H}$ are untouched by the confinement
voltage.

An important question to be addressed is the origin of the localized spins. While a model with localized spins arising from doping was suggested and analysed earlier \cite{siegert2007}, other origins can not be excluded yet. The interface between GaAs/AlGaAs could potentially induce potential fluctuations in the 2DEG. In our system the lattice constants are $a_{\rm GaAs} = 0.565325 n$m and $a_{\rm Al_{0.33}Ga_{0.67}As} = 0.56559 n$m, resulting in a lattice mismatch of $\Delta a = 0.047 \%$, and an average defect distance of $\approx 2 \mu$m. This is close to the length scale we observe, but the defects are expected to be in stripes, not in puddles, as observed in our measurements. A purely electrostatic origin for the spin lattice might also cause such ZBA, but it is difficult to imagine how the large length scale arises. Currently, experiments are carried out that will enable a more detailed insight into the origin of the localized spins \cite{Sarkozy2007}.

\section{Summary}
We report existence of an intrinsic spin lattice in high mobility
GaAs/AlGaAs heterostructures. The localized spins interact
indirectly via RKKY interaction, and the magnetic state of the
lattice can be tuned with a simple non-magnetic surface gate. The
interspin distance can be evaluated with Aharonov-Bohm like
interference at low temperatures, and commensurability effects at
higher temperatures. To probe the magnetic state directly, a Hall
geometry is introduced, and existence of fluctuations in the Hall
coefficient $\gamma_{\rm H}$ shown. The fluctuations are periodic in
$2 \delta(k_{\rm F}R) = \pi$, and deviate positively and negatively
from the non-interacting value $\gamma_{\rm H}^0$.

\section{Acknowledgment}
This project was supported by EPSRC. CS acknowledges the support of Cambridge European Trust, EPSRC, and Gottlieb Daimler and Karl Benz Foundation.

\end{document}